\documentclass[aps,pra,twocolumn,showpacs]{revtex4-1}

\usepackage{graphicx}
\usepackage{dcolumn}
\usepackage{bm}
\usepackage[version=3]{mhchem}
\usepackage{color}

\begin{document}

\title{First-principles identification of defect levels in Er-doped GaN}
\author{Khang Hoang}
\email[E-mail: ]{khang.hoang@ndsu.edu}
\affiliation{Center for Computationally Assisted Science and Technology, North Dakota State University, Fargo, North Dakota 58108, USA}

\date{\today}

\begin{abstract}

Erbium (Er) doped GaN has been studied extensively for optoelectronic applications, yet its defect physics is still not well understood. In this work, we report a first-principles hybrid density functional study of the structure, energetics, and thermodynamic transition levels of Er-related defect complexes in GaN. We discover for the first time that Er$_{\rm Ga}$-C$_{\rm N}$-$V_{\rm N}$, a defect complex of Er, a C impurity, and an N vacancy, and Er$_{\rm Ga}$-O$_{\rm N}$-$V_{\rm N}$, a complex of Er, an O impurity, and an N vacancy, form defect levels at 0.18 and 0.46 eV below the conduction band, respectively. Together with Er$_{\rm Ga}$-$V_{\rm N}$, a defect complex of Er and an N vacancy which has recently been found to produce a donor level at 0.61 eV, these defect complexes provide explanation for the Er-related defect levels observed in experiments. The role of these defects in optical excitation of the luminescent Er center is also discussed.

\end{abstract}


\maketitle


{\bf 1 Introduction} Rare-earth (RE) doped semiconductors have been of interest for optoelectronic applications, including solid-state lasers, optical fiber telecommunications, and color displays \cite{StecklMRS1999}. In the RE impurities, the partially filled $4f$-electron shell is well shielded by the more extended $5s^2$ and $5p^6$ shells, resulting in very sharp intra-$f$ optical transitions at wavelengths from the infrared to ultraviolet. GaN doped with Er, for example, emits light in a few narrow bands in the green and infrared \cite{StecklMRS1999}. The luminescent Er center can be optically excited by a direct absorption of energy into the $4f$-electron core (resonant excitation) or indirectly by transfer from the host (band-to-band excitation). In the latter mechanism, the presence of defect complexes between Er and intrinsic defects and/or other impurities in the material can play an important role. By forming defect energy levels in the host band gap that act as carrier traps, these Er-related defects can help mediate energy transfer from the host into the Er $4f$ shell. Identifying these defect levels and assigning them to specific defect configurations, however, have been challenging. 

Experimentally, Song et al.~\cite{Song2005} found several different defect levels in deep-level transient spectroscopy (DLTS) measurements, at 0.188, 0.300, 0.410, and 0.600 eV below the conduction-band minimum (CBM), in n-type GaN implanted with Er, whereas only one level, at 0.270 eV, was found in as-grown, unintentionally doped, n-type GaN. In addition, Ugolini et al.~\cite{Ugolini2007} observed a redshift of 0.19 eV in the band-edge emission of Er-doped GaN, compared to undoped GaN, and estimated the activation energy of the 1.54 $\mu$m Er emission quenching to be 191$\pm$8 meV. Their results indicate the presence of an Er-related level at about 0.2 eV below the CBM, consistent with the DLTS level at 0.188 eV reported by Song et al. This defect level has been thought to be associated with a defect complex consisting of the Er dopant and a nitrogen vacancy, Er$_{\rm Ga}$-$V_{\rm N}$. That is unlikely to be the case, however, as discussed in Ref.~\cite{HoangPSSR} and demonstrated more clearly later in our current work.

Calculations for RE impurities in GaN have been carried out by several research groups using density-functional theory (DFT) based methods, including the local-density approximation (LDA) or self-interaction corrected LDA, the generalized gradient approximation (GGA) and GGA$+U$, and LDA$+U$ within a DFT-based tight-binding approach \cite{Filhol2004,Hourahine2006,Svane2006,Mishra2007,Sanna2008,Sanna2009,Wang2012}. These studies have provided useful information on the structural and electronic properties and energetics of RE-doped GaN. However, the methods are known to have limited predictive power, especially in determining defect levels \cite{Freysoldt2014RMP}. Based on LDA calculations, for example, Filhol et al.~\cite{Filhol2004} found that RE$_{\rm Ga}$-$V_{\rm N}$ (RE = Eu, Er, Tm) possesses an energy level at about 0.2 eV below the CBM for all three REs. Their finding has often been employed to interpret the defect level at about 0.2 eV observed in experiments \cite{Song2005,Ugolini2007}. Our recent calculations based on a hybrid DFT/Hartree-Fock approach, however, showed that Er$_{\rm Ga}$-$V_{\rm N}$ instead forms an energy level at 0.61 eV below the CBM \cite{HoangPSSR}. It was also found that isolated Er$_{\rm Ga}$ is the dominant Er center and electrically inactive. We assigned the DLTS levels at 0.300 eV (as well as the 0.270 eV level observed in as-grown GaN) and 0.600 eV mentioned above to $V_{\rm N}$ and Er$_{\rm Ga}$-$V_{\rm N}$, respectively \cite{HoangPSSR}. The origins of the DLTS levels at 0.188 eV (or about 0.2 eV) and 0.410 eV are still unknown.  

In this work, we set to resolve the long-standing issue regarding the assignment of defect levels to specific defect configurations in Er-doped GaN by carrying out a study of Er-related defect complexes in wurtzite GaN using hybrid DFT/Hartree-Fock calculations. The focus is on complexes between Er, intrinsic defects, and C and O impurities. Carbon and oxygen are common unintentional dopants, although they have also been deliberately incorporated into Er-doped GaN to enhance the Er photoluminescence intensity \cite{Torvik1997,MacKenzie1998,Overberg2001}. We demonstrate that the remaining DLTS levels can be assigned to complexes consisting of Er$_{\rm Ga}$, $V_{\rm N}$, and C$_{\rm N}$ or O$_{\rm N}$. The role of these complexes in the excitation of luminescent Er centers is also discussed.

{\bf 2 Methods} Our calculations are based on DFT, using the Heyd-Scuseria-Ernzerhof (HSE) hybrid functional \cite{heyd:8207}, the projector augmented wave method \cite{PAW2}, and a plane-wave basis set, as implemented in VASP \cite{VASP2}. The Hartree-Fock mixing parameter is set to 31\%, resulting in a band gap of 3.53 eV for GaN, and the plane-wave basis-set cutoff is set to 400 eV. We model the defects using a 96-atom supercell and use a 2$\times$2$\times$2 Monkhorsk-Pack $k$-point mesh for the integrations over the Brillouin zone; spin polarization is included. All relaxations are performed with the  HSE functional and the force threshold is chosen to be 0.04 eV/{\AA}. Further details can be found in Ref.~\cite{HoangPSSR}. 

The formation energy of an intrinsic defect, impurity, or defect complex X in charge state $q$ is defined as
\begin{align}\label{eq;eform}
E^f({\mathrm{X}}^q)&=&E_{\mathrm{tot}}({\mathrm{X}}^q)-E_{\mathrm{tot}}({\mathrm{bulk}}) -\sum_{i}{n_i\mu_i} \\ %
\nonumber &&+~q(E_{\mathrm{v}}+\mu_{e})+ \Delta^q ,
\end{align}
where $E_{\mathrm{tot}}(\mathrm{X}^{q})$ and $E_{\mathrm{tot}}(\mathrm{bulk})$ are the total energies of the defect and perfect bulk supercells; $n_{i}$ is the number of atoms of species $i$ that have been added ($n_{i}>0$) or removed ($n_{i}<0$) to form the defect; $\mu_{i}$ is the atomic chemical potential. Like in Ref.~\cite{HoangPSSR}, the chemical potentials of Ga, N, Er, C, and O are referenced to the total energy per atom of bulk Ga, N$_2$ at 0 K, bulk Er, bulk C (diamond), and O$_2$ at 0 K, respectively. We consider the system in two extreme limits: Ga-rich ($\mu_{\rm Ga} = 0$) and N-rich ($\mu_{\rm N} = 0 $) conditions. Specific values of the Er, C, and O chemical potentials are determined by assuming equilibrium with ErN, bulk C, and Ga$_2$O$_{3}$ \cite{HoangPSSR}. $\mu_{e}$ is the electronic chemical potential, i.e., the Fermi level, referenced to the valence-band maximum (VBM) in the bulk ($E_{\mathrm{v}}$). $\Delta^q$ is the finite-size correction for charged defects, calculated following the procedure of Freysoldt et al.~\cite{Freysoldt}.  

The most relevant quantity for our current study is the thermodynamic transition level between charge states $q_1$ and $q_2$ of a defect, $\epsilon(q_1/q_2)$, defined as the Fermi level at which the formation energy of the defect in charge state $q_1$ is equal to that in charge state $q_2$. These levels are independent of the atomic chemical potentials and can be observed in, e.g., DLTS, where a defect in the final charge state fully relaxes to its equilibrium configuration after the transition. 

\begin{figure}[t]%
\begin{center}
\includegraphics*[width=0.88\linewidth]{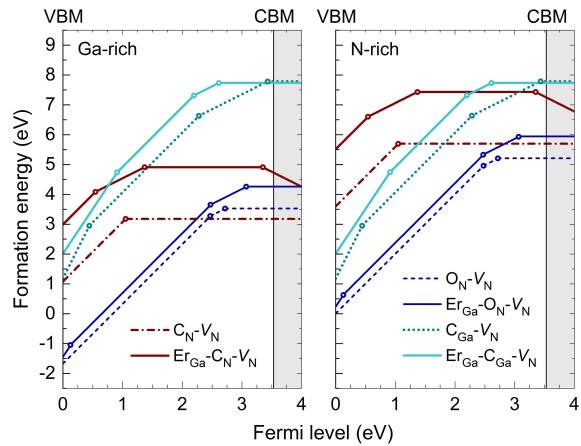}
\end{center}
\vspace{-0.5cm}
\caption{Formation energies of selected defect complexes in GaN, plotted as a function of Fermi level ($\mu_e$).  The slope in the formation energy plots indicates charge states ($q$). }
\label{fig;fe}
\end{figure}

\begin{figure*}[t]%
\begin{center}
\includegraphics*[width=0.75\linewidth]{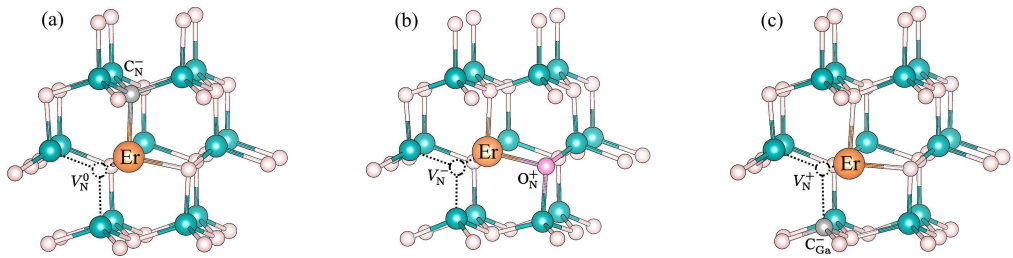}
\end{center}
\vspace{-0.5cm}
\caption{Structures of selected Er-related defect complexes in GaN: (a) (Er$_{\rm Ga}$-C$_{\rm N}$-$V_{\rm N}$)$^-$, (b) (Er$_{\rm Ga}$-O$_{\rm N}$-$V_{\rm N}$)$^0$, and (c) (Er$_{\rm Ga}$-C$_{\rm Ga}$-$V_{\rm N}$)$^0$. Large orange spheres are Er, medium blue are Ga, small gray (pink) are C (O), and smaller light pink are N.}
\label{fig;struct}
\end{figure*}

{\bf 3 Results and discussion} Figure \ref{fig;fe} shows the formation energies of selected defect complexes in GaN in their lowest-energy configurations (The results for single defects and additional complexes can be found in Ref.~\cite{HoangPSSR}). (C$_{\rm N}$-$V_{\rm N}$)$^q$ consists of a substitutional C impurity at the N site and a N vacancy. The complex has a binding energy of 0.72 eV with respect to isolated C$_{\rm N}^-$ and $V_{\rm N}^+$ when it is in charge state $q=0$ or 0.22 eV with respect to C$_{\rm N}^0$ and $V_{\rm N}^+$ in $q=+1$. C$_{\rm N}$-$V_{\rm N}$ forms one defect level in the host band gap: the (2+/0) transition level at 1.05 eV above the VBM. The complex between a substitutional O impurity at the N site and a N vacancy, O$_{\rm N}$-$V_{\rm N}$, has two defect levels: (+/0) and (2+/+) at 0.81 and 1.05 eV below the CBM, respectively. (O$_{\rm N}$-$V_{\rm N}$)$^q$ has a binding energy of 1.39 eV with respect to isolated O$_{\rm N}^+$ and $V_{\rm N}^-$ when the complex is in $q=0$; 0.23 eV with respect to  O$_{\rm N}^+$ and $V_{\rm N}^0$ in $q=+1$; or is unstable toward forming isolated O$_{\rm N}^+$ and $V_{\rm N}^+$ in $q=+2$ (the binding energy is $-0.55$ eV). Finally, C$_{\rm Ga}$-$V_{\rm N}$ consists of a substitutional C impurity at the Ga site and a N vacancy. The complex forms three defect levels in the host band gap: (+/0) and (2+/+) at 0.10 and 1.25 eV below the CBM and (4+/2+) at 0.45 eV above the VBM. (C$_{\rm Ga}$-$V_{\rm N}$)$^q$ has a binding energy of 3.45 eV with respect to isolated C$_{\rm Ga}^-$ and $V_{\rm N}^+$ when the complex is in $q=0$; 2.34 eV with respect to C$_{\rm Ga}^0$ and $V_{\rm N}^+$ in $q=+1$; or 0.80 eV with respect to C$_{\rm Ga}^+$ and $V_{\rm N}^+$ in $q=+2$. We note that as an isolated defect C$_{\rm Ga}$ is found to be stable only as C$_{\rm Ga}^+$ and act as a shallow donor, consistent with previous studies \cite{Lyons2010}.

Regarding the Er-related complexes, Er$_{\rm Ga}$-C$_{\rm N}$-$V_{\rm N}$ consists of the Er dopant, a C impurity at the axial N site, and a N vacancy at the basal N site. The Er$^{3+}$ ion is displaced from the ideal Ga site. For example, when the complex is in charge state $q=-1$ (its most stable charge state at the CBM), Er is off-center by 0.30 {\AA} toward to $V_{\rm N}$ ; see Fig.~\ref{fig;struct}(a). The two Er$-$N bonds are 2.16 {\AA} and the Er$-$C distance is 2.22 {\AA}; for comparison, the Ga$-$N bonds in the bulk are 1.95$-$1.96 {\AA}. (Er$_{\rm Ga}$-C$_{\rm N}$-$V_{\rm N}$)$^q$ has a binding energy of 0.46 eV with respect to isolated Er$_{\rm Ga}^0$, C$_{\rm N}^-$, and $V_{\rm N}^0$ (spin $S=1/2$) when the complex is in $q=-1$; 0.54 eV with respect to Er$_{\rm Ga}^0$, C$_{\rm N}^-$, and $V_{\rm N}^+$ in $q=0$; or 0.89 eV with respect to Er$_{\rm Ga}^0$, C$_{\rm N}^0$ ($S=1/2$), and $V_{\rm N}^+$ in $q=+1$. The Er$_{\rm Ga}^0$ part has a non-zero spin of $S=3/2$ if the $4f$ electrons are explicitly included in the calculations \cite{HoangPSSR}; the C$_{\rm N}^-$ and $V_{\rm N}^+$ parts both have $S=0$. In Er$_{\rm Ga}$-O$_{\rm N}$-$V_{\rm N}$, the O impurity and N vacancy are both at basal sites. The Er$^{3+}$ ion is off-center by 0.26 {\AA} toward $V_{\rm N}$ when the complex is in $q=0$ ; see Fig.~\ref{fig;struct}(b). The two Er$-$N bonds are 2.17 and 2.21 {\AA} and the Er$-$O distance is 2.21 {\AA}. (Er$_{\rm Ga}$-O$_{\rm N}$-$V_{\rm N}$)$^q$ has a binding energy of 2.22 eV with respect to isolated Er$_{\rm Ga}^0$, O$_{\rm N}^+$, and $V_{\rm N}^-$ ($S=1$) when the complex is in $q=0$; 1.40 eV with respect to Er$_{\rm Ga}^0$, O$_{\rm N}^+$, and $V_{\rm N}^0$ ($S=1/2$) in $q=+1$; or 0.64 eV with respect to Er$_{\rm Ga}^0$, O$_{\rm N}^+$, and $V_{\rm N}^+$ in $q=+2$. Finally, Er$_{\rm Ga}$-C$_{\rm Ga}$-$V_{\rm N}$ consists of the Er dopant, a C impurity at the Ga site, and a N vacancy at the basal N site. In the neutral charge state, the Er$^{3+}$ ion of the complex is off-center by 0.62 {\AA} toward $V_{\rm N}$ and C$_{\rm Ga}$; see Fig.~\ref{fig;struct}(c). (Er$_{\rm Ga}$-C$_{\rm Ga}$-$V_{\rm N}$)$^q$ has a binding energy of 5.05 eV with respect to Er$_{\rm Ga}^0$, C$_{\rm Ga}^-$, and $V_{\rm N}^+$ when the complex is in $q=0$; 2.34 eV with respect to Er$_{\rm Ga}^0$, C$_{\rm Ga}^0$, and $V_{\rm N}^+$ in $q=+1$; or 1.50 eV with respect to Er$_{\rm Ga}^0$, C$_{\rm Ga}^+$, and $V_{\rm N}^+$ in $q=+2$. In certain charge states, there is thus charge transfer between the non-Er constituents forming the complexes. The identity of the constituent defects is largely preserved. The electronic behavior of the complexes is, however, significantly different from that of the isolated defects, due to the strong local elastic and electrostatic interaction between the constituents. Unlike the isolated Er$_{\rm Ga}$ which has been found to be electrically inactive and induce only a small lattice distortion \cite{HoangPSSR,Filhol2004,Hourahine2006,Mishra2007,Sanna2008,Sanna2009}, the Er-related complexes thus introduce defect levels in the host band gap (see Fig.~\ref{fig;fe}) and have the Er center displaced significantly from the ideal lattice site. We note that the defect configuration presented in Fig.~\ref{fig;struct}(a) is 0.24 eV lower in energy than one where C$_{\rm N}^-$ is at the basal site and $V_{\rm N}^0$ at the axial site and 0.04 eV lower than one where the defects are both at basal sites. The defect configuration in Fig.~\ref{fig;struct}(b) is 0.05 eV (0.07 eV) lower than one where O$_{\rm N}^+$ is at the axial (basal) site and $V_{\rm N}^-$ at the basal (axial) site. Other (Er$_{\rm Ga}$-C$_{\rm Ga}$-$V_{\rm N}$)$^0$ configurations are $>$1.3 eV higher in energy than the one in Fig.~\ref{fig;struct}(c).

\begin{figure}[t]%
\begin{center}
\includegraphics*[width=0.88\linewidth]{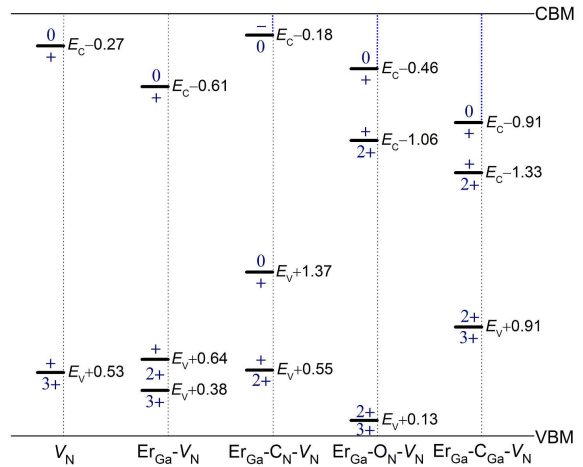}
\end{center}
\vspace{-0.4cm}
\caption{Thermodynamic transition levels associated with selected intrinsic defects and Er-related defect complexes in GaN; $E_{\rm V}$ and $E_{\rm C}$ are the energies at the VBM and CBM, respectively.}
\label{fig;levels}
\end{figure}

In n-type GaN and under Ga-rich condition, the formation energies of Er$_{\rm Ga}$-C$_{\rm N}$-$V_{\rm N}$, Er$_{\rm Ga}$-O$_{\rm N}$-$V_{\rm N}$, and Er$_{\rm Ga}$-C$_{\rm Ga}$-$V_{\rm N}$ are 4.73, 4.26, and 7.73 eV, respectively, with respect to the chosen set of the atomic chemical potentials. For comparison, the formation energy of Er$_{\rm Ga}$ is 1.55 eV and that of Er$_{\rm Ga}$-$V_{\rm N}$ is 3.82 eV \cite{HoangPSSR}. The formation energies of the Er-related complexes are thus relatively high. However, it is expected that the complexes can still occur with a non-negligible concentration, especially under non-equilibrium condition where the defect concentration is not directly determined by the formation energy. Besides, the formation energies can be much lower at low Fermi-level values; see Fig.~\ref{fig;fe}. Also under such a non-equilibrium condition, constituents of a defect complex can be kinetically trapped, making the complex stable even when it has a relatively low binding energy, such as in the case of Er$_{\rm Ga}$-C$_{\rm N}$-$V_{\rm N}$. 

Figure~\ref{fig;levels} shows defect energy levels formed by the Er-related defect complexes. The results for $V_{\rm N}$ and Er$_{\rm Ga}$-$V_{\rm N}$ have been reported in Ref.~\cite{HoangPSSR} but are also included here for completeness. Among all plausible single point defects and defect complexes explored thus far, we find that only Er$_{\rm Ga}$-C$_{\rm N}$-$V_{\rm N}$ produces a defect level at approximately 0.2 eV below the CBM. In light of our results, we assign the DLTS levels in Er-implanted GaN reported by Song et al.~\cite{Song2005} to the following defect configurations: the 0.188 eV level (and the defect level at about 0.2 eV reported in Ref.~\cite{Ugolini2007}) to the (0/$-$) level at 0.18 eV of Er$_{\rm Ga}$-C$_{\rm N}$-$V_{\rm N}$, the 0.300 eV level (and the DLTS level at 0.270 eV in as-grown GaN) to the (+/0) level at 0.27 eV of $V_{\rm N}$, the 0.410 level to the (+/0) level at 0.46 eV of Er$_{\rm Ga}$-O$_{\rm N}$-$V_{\rm N}$, and the 0.600 eV level to the (+/0) level at 0.61 eV of Er$_{\rm Ga}$-$V_{\rm N}$. 

The deep defect levels associated with the Er-related complexes are expected to act as carrier traps; an electron captured here can subsequently recombine non-radiatively with a free hole from the valence band or a hole at some acceptor level and transfer energy to the Er $4f$-electron shell. Also, due to the attractive Coulomb interaction, it is expected that the electron will be efficiently trapped by the positively charged constituent in the electron-capturing defect configurations; i.e., the $V_{\rm N}^+$ part of (Er$_{\rm Ga}$-$V_{\rm N}$)$^+$, (Er$_{\rm Ga}$-C$_{\rm N}$-$V_{\rm N}$)$^0$, and (Er$_{\rm Ga}$-C$_{\rm Ga}$-$V_{\rm N}$)$^+$ or the O$_{\rm N}^+$ part of (Er$_{\rm Ga}$-O$_{\rm N}$-$V_{\rm N}$)$^+$. The Er-related defect complexes thus have different electronic behavior and provide different local environments for the Er dopant which can lead to multiple optically active Er centers as observed in experiments \cite{Kim1997,Przybylinska2001,Braud2003,Dierolf2004,GeorgeAPL2015}. The significant local lattice distortion around the Er centers helps relax the selection rules and allows for bright emission. Our results also support experimental data showing direct evidence of trap-mediated excitation in Er-doped GaN \cite{Bodiou2008} and provide explanation for the enhanced Er photoluminescence intensity observed in GaN co-doped with Er and C and/or O \cite{Torvik1997,MacKenzie1998,Overberg2001}. 

{\bf 4 Summary and conclusions} We have found that Er-related defect complexes form energy levels in the host band gap. The defect levels observed in DLTS and excitation dynamics studies of Er-doped GaN, including the one at about 0.2 eV below the CBM often mentioned in the literature, can be attributed to specific defect configurations that involve Er, intrinsic defects, and (intentional or otherwise) C or O dopants: Er$_{\rm Ga}$-$V_{\rm N}$, Er$_{\rm Ga}$-C$_{\rm N}$-$V_{\rm N}$, and Er$_{\rm Ga}$-O$_{\rm N}$-$V_{\rm N}$. Although these defect complexes are not the dominant Er center (the isolated Er$_{\rm Ga}^0$ is), they can be efficient defect-related Er centers for band-to-band excitation.

{\bf Acknowledgements} This work was supported by the U.S. Department of Energy Grant No.~DE-SC0001717 and by the Center for Computationally Assisted Science and Technology (CCAST) at North Dakota State University. 

\providecommand{\WileyBibTextsc}{}
\let\textsc\WileyBibTextsc
\providecommand{\othercit}{}
\providecommand{\jr}[1]{#1}
\providecommand{\etal}{~et~al.}

\end{document}